\documentclass[twocolumn,aps,prb,showpacs,floatfix]{revtex4}

\usepackage{graphicx}%
\usepackage{dcolumn}
\usepackage{bm}%

\def\dfrac#1#2{{\displaystyle{#1\over#2}}}

\begin{document}
\title{Effect of the Surface on the Electron Quantum Size Levels
and Electron $g$-Factor in Spherical Semiconductor Nanocrystals}

\author{A.V. Rodina}
\thanks{On leave from A. F. Ioffe Physico-Technical
Institute, 194021, St.-Petersburg, Russia.} \affiliation{Institute
for Solid State Physics, TU Berlin, D-10623 Berlin, Germany}
\author{Al. L. Efros}
\affiliation{Naval Research
Laboratory, Washington, DC 20375, USA}
\author{A.~Yu.~Alekseev}
\affiliation{Institute of Theoretical Physics, Uppsala University,
S-75108, Uppsala, Sweden \\
and Department of Mathematics, University of Geneva, 1211
Geneva, Switzerland}
\date{\today}

\begin{abstract}
The structure of the electron quantum size levels
in spherical nanocrystals is studied in the framework of an eight--band
 effective mass model at zero and weak magnetic fields.  The effect
of the nanocrystal surface is modeled through the boundary condition
imposed on the envelope wave function at the surface. We show that
the spin--orbit splitting of the valence band leads to the
surface--induced spin--orbit splitting of the excited conduction band
states and to the additional surface--induced magnetic moment for
electrons in bare nanocrystals. This additional magnetic moment
manifests itself  in a nonzero surface contribution to the linear
Zeeman splitting of all quantum size energy levels including the
ground $1S$ electron state. The fitting of the size dependence of the
ground state electron $g$ factor in CdSe nanocrystals has allowed us
to determine the appropriate surface parameter of the boundary
conditions. The structure of the  excited electron states  is
considered in the limits of weak  and strong magnetic fields.
\end{abstract}

\maketitle

\section{Introduction}

 The modern technique of nanocrystal growth has created high optical
quality nanocrystals with narrow size distribution\cite{brus,Murray}
that can be also chemically doped.\cite{Sionest} This allows one to
study the level structure  of the nanocrystals using interband
optical spectroscopy,\cite{josa,norris,Klimov} tunneling
spectroscopy,\cite{banin,millo} and the far--infrared intraband
spectroscopy.\cite{Sionest,shim} Important additional information
about the origin of these levels and their symmetry  could be
obtained by using optical magnetospectroscopy of the intraband and
interband transitions and tunnelling magnetospectroscopy,  as was
done many years ago in the case of atoms, hydrogenlike shallow
defects and excitons. There is the other group of experiments, such
as magneto circular dichroism,\cite{kuno} spin--flip Raman
scattering,\cite{sirenko} time dependent Faraday
rotation,\cite{gupta,gupta02}  EPR  and ODMR
measurements,\cite{nozik1,nozik2,lifshitz}  and
 magneto-photoluminescence,\cite{Efrosprb96} where the magnetic field
is an essential component of the experimental technique.
Inrepretation of all these experiments requires the knowledge of
electron and hole energy spectra  in magnetic fields.

A wide class of semiconductor materials can be prepared in
nanocrystal form. In most cases they have  a spherical shape and can
be  bare semiconductor nanocrystals or  onion type heterostructure
nanocrystals composed of an inner semiconductor core coated with
several spherical shells of different
semiconductors.\cite{alivisatos,weller,bawendi} Technology allows one to
vary the radius of the nanocrystal core from 10--40\,\AA\ in a
controlled manner. The energy spectra of the NCs are formed by the
discrete quantum size levels (QSLs). The QSLs of the
confined electrons in the conduction band are characterized by the value
of the orbital momentum $l$ and the total angular momentum $j=l \pm
1/2$, similar to the bound electron states in atoms\cite{bethe}. In
a zero magnetic field, the levels are degenerate with respect to the
projection of the total momentum $j$.

The spectra of the nanocrystals in an external magnetic field $H$ is
well described as an atomic Zeeman effect.  A diamagnetic $H^2$
contribution to the spectra could be neglected because the
nanocrystal radius, $a$, is considerably smaller than the magnetic
length, $L=\sqrt{e\hbar/cH}$, where  $e$ is the absolute value  of a
free electron charge, $c$ is the speed of light. In a reasonable
magnetic field $H<10$\,T the diamagnetic contribution to the QSLs
$\sim (a/L)^4\ll1$. However, the effect of the external magnetic
field on the electron QSLs depends on the zero field spin-orbit
splitting, $\Delta_c$, of the levels with different $j = l \pm 1/2$
(One can find a description of the similar effect for atoms in Ref.
\onlinecite{bethe}). If $\Delta_c$ is larger than the Zeeman energy,
the weak magnetic field splits the electron levels:
\begin{equation}
\Delta E_j^{\pm}=\mu_B g_j^{\pm} m H~, ~~~m=-j,-j+1,...,j~,
\end{equation}
where $\mu _{B}=e\,\hbar /2m_{0}\,c$ is the Bohr magneton, $m_0$ is
the free electron mass, $m$ is the projection of the total  momentum
on the magnetic filed direction and $g_j^{\pm}$ is the effective
$g$-factor of the corresponding state, that is an analog of the {\em
Land}$\acute{e}$ \, {\em factor}\, for atoms.   If $\Delta_c$ is
small, the Zeeman splitting can be described as the sum of the spin
and orbital contributions:
\begin{equation}
\Delta E_l=\mu_Bg_s s_zH+ \mu_B g_l l_z H~,
\label{delsl}
\end{equation}
where $s_z=\pm 1/2$ and $l_z=-l,-l+1,...,l$  are the electron spin
and angular momentum projections on a magnetic field direction
correspondingly, and $g_s$ and $g_l$ are the spin and orbital
g-factors of the electron in the  corresponding state.  In bare
nanocrystals made of semiconductors with a simple parabolic
conduction band $g_s$ is equal to the bulk  electron effective
$g$--factor $g_c$ and $g_l=m_0/m_c$, where $m_c$ is the bulk electron
effective mass.

Spin--flip Raman scattering studies of CdS
nanocrystals,\cite{sirenko} however,  have demonstrated the
dependence of the ground 1S electron state $g_s$ on the excitation
energy, and thus on the energy of the electron state. The variation
of $g_s$ with energy can be connected with the energy dependence of
the bulk electron $g$--factor.\cite{sirenko}   The eight band Kane
model, that is used to describe the electron energy spectra in most
direct gap semiconductors, gives the following  dependence of the
electron $g$--factor $g_c(E)$ on its energy, $E$:\cite{roth,weisbuch}
  \begin{equation}
 g_c(E)=g^* -{2E_p\over 3}{\Delta \over (E_g+E)(E_g+\Delta+E)}~.
\label{bulkg}
\end{equation}
Here  $E_g$ is the band gap energy, $\Delta$ is  the spin--orbit
splitting of the top of the valence band, $E_p$ is the energetic Kane
parameter, and  $g^*=g_0 + \Delta g$, where $g_0=2$  is the free
electron $g$--factor and    $|\Delta g| \ll 1 $  is the contribution
of remote bands to the  bulk  electron  $g$ factor. The second term
in Eq. (\ref{bulkg}) describes the negative contribution of the
valence band (with $\Delta > 0$) into the electron $g$-factor. It
decreases with increasing the energy of QSLs caused by the reduction
of the nanocrystals size. Thus, the energy dependence of $g_c(E)$ in
Eq. (\ref{bulkg}) is conducive to  the size dependence  of $g_s$ in
Eq. (\ref{delsl}), that has been measured recently in bare and
core-shell CdSe nanocrystals.\cite{gupta02}

The size dependence  of $g_s$ for the ground 1S electron state in the
spherical heterostructure has been calculated  in Ref.
\onlinecite{kiselev}  within the eight band  Kane model. It has been
shown that  in a spherical heterostructure formed by two
semiconductors, $A$ and $B$, $g_s$  is the sum of the  weighted
volume contributions of each material and the  interface
contributions: $g_s= \overline{g_{s}(A)}+\overline{g_{s}(B)}+
g_{AB}$. The interface term, $g_{AB}$, is proportional to the square
of the conduction band component of the wave function, $f_c$, at the
$A/B$ heterointerface. This calculation   shows clearly that $g_s$ is
very sensitive to the value of the wave function at the
heterointerface and to its leakage under the barrier and thus to the
boundary conditions (BCs) imposed on the wave function at the
heterointerface.  In Ref. \onlinecite{kiselev}  the size dependence
of the electron g-factor is calculated using standard BCs that
assumes a continuity of $f_c$ at the interface.   This leads to the
vanishing of the interface term, $g_{AB}$, in the bare semiconductor
nanocrystals, that are modeled by the infinite potential barrier in
the layer $B$  because the standard  BC at the surface in this case
is $f_c=0$.

However, the  standard BCs are not always justified even for infinite
potential barriers (see for example Refs. \onlinecite{brag99,bc}). In
general, the wave function $f_c$ does not  vanish  at the surface of
bare semiconductor nanocrystals and satisfies the general  boundary
conditions that are the characteristics of the particular
surface.\cite{brag99,bc} The effect of a semiconductor surface on the
light absorption in indirect band  semiconductor nanocrystals has
been studied in Ref. \onlinecite{brag99} and was found to be
significant.

In this paper we study the effect of the general boundary conditions
on the QSLs of an electron in zero and  weak external magnetic
fields. We show that there is an additional surface--induced
spin--orbit term for the electron states whose magnitude is
proportional to $\Delta |f_c(a)|^2\, a^3$.  This surface--induced
spin--orbit interaction leads the additional magnetic moment of the
electrons in an external magnetic field  in analogy with the
additional relativistic magnetic moment of the electrons in
atoms.\cite{abragam,margenau}  This is conducive to the nonzero
surface contribution to the electron $g$ factor that controls the
linear Zeeman splitting of the QSLs. The fitting of the experimental
size dependence of the electron $g$ factor in the ground
state\cite{gupta02} has allowed us to determine the appropriate
parameter of the general boundary conditions for bare CdSe
nanocrystals. The spectra of excited electron states has been
considered in the two limits of the relation between the zero field
spin--orbit splitting and the electron energy in magnetic field
similar to consideration conducted for atoms.\cite{bethe}

The paper is organized in the following way: In Sec. \ref{theory}  we
describe the eight--band Hamiltonians that include the external
magnetic field effects. In Sec. \ref{spher} we derive the general
boundary conditions for the conduction band component of the envelope
function at the abrupt surface (see Sec. \ref{spherbc}) and analyze
the surface effect on the electron QSLs in a zero magnetic field (see
Sec. \ref{spherso}). The effect of the weak external magnetic field
on the electron QSLs is considered in Sec.  \ref{magnetic}. The size
dependence of the electron ground state $g$ factor  in bare CdSe
nanocrystals has been calculated and the value of the
 surface boundary parameter has been determined  by fitting experimental
data in Sec. \ref{ground}. The symmetry of the electron excited states in
an external magnetic field is studied in the low and strong magnetic field
regimes (see Sec. \ref{excited}). The results are summarized and
discussed in Sec. \ref{sum}.

\section{\label{theory}Effect of a weak magnetic field on the QSLs
 within the eight--band model: perturbation approach}

\subsection{The eight band model}
The energy band structure of the cubic semiconductors near the
center of the first Brillouin zone can be well described within the
eight--band $\mathbf{k \cdot p}$  model.\cite{pb,bir,ivchenko} In
the homogenous semiconductor, the  full wave function  can be
presented as the expansion:\cite{bir}
\begin{eqnarray}
{\Phi (\mbox{\boldmath $ r$})} &=& \sum_{\mu = \pm 1/2}
\Psi^{c}_\mu(\mbox{\boldmath $ r$}) |S\rangle
u_{\mu} \, +\nonumber \\
&\, &\sum_{ \mu = \pm 1/2  } \sum_{\alpha=x,z,z}  \Psi_{
\alpha\,\mu}^v(\mbox{\boldmath $ r$}) |R_\alpha \rangle u_{\mu} \,
,\label{psi}
\end{eqnarray}
where  $u_{1/2}$ and $u_{-1/2}$ are the eigenfunctions of the spin
operator $  \mbox{\boldmath $\hat S$} = 1/2 \, \mbox{\boldmath $\hat
\sigma$}$, where $\mbox{\boldmath $ \hat \sigma$}= \{ \hat \sigma_x ,
\hat \sigma_y ,  \hat \sigma_z \}$  are the Pauli matrices,
$|S\rangle $ is the Bloch function of the conduction band edge at the
$\Gamma$--point of Brilluoin zone representing the eigenfunction of
internal momentum $I=0$, and $|R_x\rangle=|X\rangle$,
$|R_y\rangle=|Y\rangle$, $|R_z\rangle=|Z\rangle$ are the Bloch
functions of the valence band edge at the $\Gamma$--point of
Brillouin zone. The combination of these functions:
$1/\sqrt{2}(|R_x\rangle \pm i |R_y\rangle)$ and $|R_z\rangle$  are
the eigenfunctions of the internal momentum $I=1$ with projections
$\pm 1$ and $0$ on the $z$ axis respectively (see Refs.
\onlinecite{bir,lut}).  The smooth functions $\Psi^{c}_{\pm
1/2}(\mbox{\boldmath $r$})$ are the components of the conduction band
spinor envelope function $\Psi^c=\left( \begin{array}{c} \Psi^c_{1/2}
\\ \Psi^c_{-1/2} \end{array} \right)$, and $\Psi^{v}_{x\,\pm
1/2}(\mbox{\boldmath $r$})$, $\Psi^{v}_{y\,\pm 1/2}(\mbox{\boldmath
$r$})$, $\Psi^{v}_{z\,\pm 1/2}(\mbox{\boldmath $r$})$ are the $x,y,z$
components of the valence  band spinor envelope vector
${\bf \Psi^v}=\left( \begin{array}{c} {\bf \Psi^v}_{1/2} \\
 {\bf \Psi^v}_{-1/2} \end{array} \right)=\left\{ \left(
\begin{array}{c} \Psi^v_{x\,1/2} \\ \Psi^v_{x\,-1/2} \end{array}
\right),
\left( \begin{array}{c} \Psi^v_{y\,1/2} \\
\Psi^v_{y\,-1/2} \end{array} \right),
\left( \begin{array}{c}
\Psi^v_{z\,1/2} \\ \Psi^c_{z\,-1/2} \end{array} \right)\right\}$.
In the presence of external magnetic field the eight--component
envelope function ${\Psi}(\mbox{\boldmath $r$}) \equiv \{
{\Psi}^c(\mbox{\boldmath $ r$}), {\bf \Psi}^v(\mbox{\boldmath $
r$})\}$ is the solution of the Schr\"{o}dinger equation\cite{pb}
\begin{eqnarray}
\hat{H}(  \mbox{\boldmath $\hat k$})\left( \begin{array}{c}
 \Psi^c \\ {\bf \Psi}^v \end{array} \right) =
  E \left(\begin{array}{c} \Psi^c \\
   {\bf \Psi}^v \end{array} \right)\, ,\qquad \nonumber \\
    \hat{H}( \mbox{\boldmath $\hat  k$})=
    \left(  \begin{array}{cc} \hat{H}^c(\mbox{\boldmath $\hat  k$})&
     i \hbar P \hat U_2 \mbox{\boldmath $\hat  k$} \\
- i \hbar P \hat U_2 \mbox{\boldmath $\hat  k$} &
\hat{H}^v( \mbox{\boldmath $\hat  k$})
\end{array} \right)
 \, \,  . \label{hfull}
\end{eqnarray}
Here,  the energy $E$ is measured from the bottom of the conduction
band,  $ \mbox{\boldmath $ \hat k$}=\dfrac{1}{\hbar} \left(
\mbox{\boldmath $ \hat p$}+\dfrac{e}{c}\mbox{\boldmath $ A$} \right)
$, where $ \mbox{\boldmath $\hat  p$}= - i\hbar \mbox{\boldmath $
\nabla$} $ is the momentum operator and  $\mbox{\boldmath $
A$}=(1/2)[\mbox{\boldmath $ H$}\times {\mbox{\boldmath $ r$}}]$ is
the vector potential of the magnetic field.  The $2\times 2$ unit
matrix, $\hat U_2$, and  the Pauli matrices, $\sigma_x$, $\sigma_y$
and $\sigma_z$, are  acting on the spinor components of the wave
functions ($\mu= \pm 1/2$). $P= -i\langle S|\hat{p}_z|Z\rangle/m_0$
is  the Kane matrix element describing the coupling of the conduction
and valence bands (The Kane energy parameter: $E_p=2m_0 P^2$, was
introduced in Eq. (\ref{bulkg}),  respectively).   The   off-diagonal
matrix element in Eq. (\ref{hfull}) acts on the ${\bf \Psi}^v$ as a
scalar product $\sim \mbox{\boldmath $\hat  k$}{\bf
\Psi}^v\equiv\hat{k}_x \Psi^v_x+\hat{k}_y \Psi^v_y+\hat{k}_z
\Psi^v_z$.  The conduction band part of the Hamiltonian, $\hat{H}^c$,
acting on the spinor function $\Psi^c$ has  the form:
\begin{equation}
\hat{H}^c(\mbox{\boldmath $\hat  k$})= \frac{\alpha \hbar
^{2}}{2m_{0}}\hat U_2  \hat k^{2} + \frac{1}{2} g^* \mu _{B}(
\mbox{\boldmath $\hat  \sigma H$}), \label{h2}
\end{equation}
where  $g^*$ is defined in Eq. (\ref{bulkg}),  and $\alpha$ takes
into  account the contribution  of remote bands   to  the
energy--dependent electron effective mass, $m_c(E)$. In  cubic and
zinc-blende semiconductors this dependence has the following form:
\begin{eqnarray}
{m_0\over m_c(E)}=\alpha+{E_p\over 3} \left[{2\over E_g + E} +{1\over
E_g+E+\Delta}\right] \, .
\label{bulkm}
\end{eqnarray}
 In this paper we  focus only on the electron QSLs with energies
$0<E<Eg$. This allows us to neglect hereafter the $k^2$ terms in the
valence band part of the Hamiltonian, $\hat{H}^v$, and to present it
in the spherical approximation  as:\cite{bir,lut}
\begin{eqnarray}
\hat{H}^v( \mbox{\boldmath $\hat  k$})&=&-\left( E_g +
\frac{1}{3}\Delta + (1+3 \kappa )\,({\mbox{\boldmath $\hat
I$}}\mbox{\boldmath $ H$}) \right) \hat U_2 \,   \nonumber \\ &+&\,
\frac{1}{3}\Delta \, (\mbox{\boldmath $\hat  I$}\mbox{\boldmath $\hat
\sigma$})+ \frac{1}{2} \mu _{B}g_{0} ( \mbox{\boldmath $\hat \sigma$}
\mbox{\boldmath $ H$}). \label{h6}
\end{eqnarray}
 Here, the  Hamiltonian $\hat H^v$ should be considered as the $2
\times 2$ matrix   acting on the spinor vector ${\bf \Psi}^v$ rather
than   the $6 \times 6$ matrix Hamiltonian acting on the six
component wave function as in Ref. \onlinecite{lut}. Correspondingly,
$\mbox{\boldmath $\hat I$}=\{\hat I_x, \hat I_y, \hat I_z \}$ in Eq.
(\ref{h6}) is the vector operator. It is straightforward to show that
in this representation $(\mbox{\boldmath $\hat  I$}\mbox{\boldmath $
T$}){\bf \Psi}^v = - i [\mbox{\boldmath $ T$} \times {\bf \Psi}^v]$,
where $\mbox{\boldmath $ T$}$ is an arbitrary vector. The terms
proportional to $\Delta$ in Hamiltonian (\ref{h6})   describe the
effect of the spin--orbit coupling  and
$\kappa=\kappa^L-{E_p/6E_g}$,\cite{pb} where   $\kappa^L$ is the
magnetic Luttinger parameter.\cite{lut}  One can find that $\kappa
=-2/3$  from the relation:\cite{pb,dkk} $\kappa^L= (-2 - \gamma_1^L +
2\gamma_2^L+3\gamma_3^L)/3$,  where  the  effective mass Luttinger
parameters,\cite{lut} $\gamma_{1,2,3}^L$, can be presented as:
$\gamma_1^L\approx {E_p/3E_g}$ and $\gamma_{2}^L\approx
\gamma_{3}^L\approx {E_p/6E_g}$ in the approximation used
above.\cite{pb}

\subsection{Linear Zeeman effect as a perturbation}

Following the perturbation approach developed in Ref.
\onlinecite{kiselev}, we present the Hamiltonian (\ref{hfull}) as:
\begin{equation}
\hat H_{PB}(\mbox{\boldmath $\hat  k$})= \hat H^0 (\mbox{\boldmath
$\hat  p$})+ \hat H^{'}(\mbox{\boldmath $\hat  p$}, \mbox{\boldmath $
H$}) \, \, , \label{hpert}
\end{equation}
where the Hamiltonian $\hat H^0$ describes  the
electron  states in  zero magnetic field:
\begin{eqnarray}
\hat{H}^0( \mbox{\boldmath $\hat  p$}) = \left( \begin{array}{cc}
  \frac{\alpha }{2m_{0}}\hat U_2\hat p^{2}  &
  i  P \hat U_2 \mbox{\boldmath $\hat  p$} \\
\, - i P \hat U_2 \mbox{\boldmath $\hat  p$} &
  \,\frac{1}{3}\Delta \,(\mbox{\boldmath $\hat  I \hat  \sigma$})-
  \left( E_g + \frac{1}{3}\Delta \right)\hat U_2
\end{array} \right),~
\label{h0}
\end{eqnarray}
and the Hamiltonians $ \hat H^{'}$ describes the effect of a weak
external magnetic field.   Linear in the magnetic field terms of
$\hat H^{'}(\mbox{\boldmath $\hat  p$}, \mbox{\boldmath $H$})$  can
be written:
\begin{equation}
\hat H^{'}(\mbox{\boldmath $\hat  p$}, {H}) = \mu_{B} H \hat G +
\frac{e}{c} \{ \mbox{\boldmath $ A$}, \mbox{\boldmath $\hat  V$}
\}\equiv \mu_{B} H ( \hat G  + \hat G^{'} ),
\end{equation}
 where  $\{a,b\} = 1/2( ab +  ba )  $   denotes an anticommutator and
$\mbox{\boldmath $\hat  V$} = \partial \, \hat H^0 (\mbox{\boldmath $
p$}) \, / \, \partial \, \mbox{\boldmath $p$}$ is the envelope
velocity operator. The operators $\hat G$ and $\hat G^{'}$ are
defined as:
\begin{eqnarray}
\hat G = \left(
\begin{array}{cc}
\frac{1}{2} g^{*} (\mbox{\boldmath $\hat \sigma$} \mbox{\boldmath $ n$})  & 0\\
 0 & (\mbox{\boldmath $\hat  I$}\mbox{\boldmath $ n$})\hat U_2+
 \frac{1}{2} g_{0} (\mbox{\boldmath $\hat
\sigma$} \mbox{\boldmath $ n$})
\end{array} \right) \, ,
\end{eqnarray}
\begin{eqnarray}
\hat G^{'} =
\left(
\begin{array}{cc}
\alpha (\mbox{\boldmath $\hat  L$} \mbox{\boldmath $ n$}) &
   i P  \dfrac{m_0}{\hbar} [\mbox{\boldmath $ n$} \times \mbox{\boldmath $ r$}]\\
-i P  \dfrac{m_0}{\hbar}  [\mbox{\boldmath $ n$} \times \mbox{\boldmath $ r$}]\,& 0
\end{array} \right)\hat U_2\, ,
\end{eqnarray}
where $\mbox{\boldmath $ n$} = \mbox{\boldmath $ H$}/H$  is the unit
vector directed along the external magnetic field, and
$\mbox{\boldmath $\hat  L$} = \frac{1}{\hbar} \left[ \mbox{\boldmath
$ r$} \times \mbox{\boldmath $\hat  p$} \right]$ is the envelope
angular momentum operator.

 Generally, the calculation  of the linear Zeeman effect for the
carriers confined in a $3D$ external potential of arbitrary shape is
a straightforward theoretical problem. The same procedure is used for
an impurity potential and for a zero dimensional quantum dot of an
arbitrary shape. Using the unperturbed Hamiltonian $\hat H^0$ one has
to find first the energy levels $E_\lambda$ and the corresponding
eigen functions
 $\Psi_\lambda=\{\Psi^c,{\bf \Psi}^v\}$ of the carrier in the external confining
 potential. Then, the effect of the magnetic field can be found perturbatively:
\begin{equation}
\Delta E_\lambda = \mu_B H\left[ \left< {\Psi}_{\lambda} | \hat G |
{ \Psi}_{\lambda} \right> + \left< {\Psi}_{\lambda} | \hat G^{'} |
{\Psi}_{\lambda} \right> \right] .
\end{equation}
where the wave functions should be normalized $\left<
{\Psi}_{\lambda} |  { \Psi}_{\lambda} \right>=1$.  Substituting
$\Psi_\lambda$ into this expression we obtain:
\begin{equation}
\Delta E_\lambda = \Delta E_c + \Delta E_v + \Delta E_{coup} \, ,
\end{equation}
where the contribution of the valence and conduction bands to Zeeman
effect $\Delta E_c$ and $\Delta E_v$ correspondingly are:
\begin{eqnarray}
\frac{\Delta E_c}{\mu_B H}= \frac{1}{2} \left< {\Psi}^c |
 g^* (\mbox{\boldmath $\hat \sigma$} \mbox{\boldmath $n$}) |
 { \Psi}^c \right> +
 \left< {\Psi}^c_\mu |\alpha (\mbox{\boldmath $\hat  L$} \mbox{\boldmath $ n$})
  |  { \Psi}^c_\mu \right> \, ,
\label{dec} \\
\frac{\Delta E_v}{\mu_B H}= \frac{1}{2} \left< {\bf \Psi}^v |
 g_0 (\mbox{\boldmath $\hat \sigma$} \mbox{\boldmath $n$}) |
  {\bf \Psi}^v \right> +
 \left< {\bf \Psi}^v_\mu |( \mbox{\boldmath $\hat  I$} \mbox{\boldmath $ n$}) |
  {\bf \Psi}^v_\mu \right> \, ,
\label{dev}
\end{eqnarray}
and the coupling correction is given by
\begin{eqnarray}
\frac{\Delta E_{coup}}{\mu_B H } &= &   \left< {\Psi}^c | i
\frac{m_0P}{\hbar} \hat U_2 [\mbox{\boldmath $ n$}\times
\mbox{\boldmath $ r$}] |  { \bf \Psi}^v \right> \nonumber \\
&+&\left< {\Psi}^c | i \frac{m_0P}{\hbar}\hat U_2 [\mbox{\boldmath $
n$}\times \mbox{\boldmath $ r$}] |  { \bf \Psi}^v \right>^*   \, .
\label{dcoup}
\end{eqnarray}
 In principle, this straightforward approach allows us to find the
 linear Zeeman splitting of the energy levels in any arbitrarily
 shaped hetero-nanostructure. However, the calculations of the zero
 field wave functions $\Psi_\lambda$ for arbitrary shape
 heterostructure is a rather cumbersome procedure and, consequently,
 the  Zeeman splitting calculation that uses these wave functions is
 very complicated.  In the present paper, from hereafter, only
 spherical semiconductor nanocrystals are considered. The high
 symmetry of these quantum dots allow us to calculate the Zeeman
 splitting of the QSLs.

\section{\label{spher}Energy levels of the electron confined in bare spherical nanocrystals }

\subsection{\label{spherbc}General boundary conditions for the envelope wave functions}

We will consider bare semiconductor nanocrystals which surface can be
modeled by an impenetrable barrier.  To find unperturbed envelope
wave functions that are described by the Hamiltonians of  Eq.
(\ref{h0}) in the bulk region one should impose  an appropriate
boundary conditions on these functions at the nanocrystal surface.
For the impenetrable barrier the general boundary conditions (GBCs)
have been shown in Ref. \onlinecite{bc}  to guarantee the vanishing
of the normal to the surface component of the envelope flux density
matrix. In a spherical nanocrystal the normal envelope flux density
matrix, $J_{\tau}^{\lambda  \, \eta}(\mbox{\boldmath $r$})$,  should
vanish at any point of the nanocrystals surface:
\begin{eqnarray}
J_{\tau}^{\lambda  \, \eta}(\mbox{\boldmath $r$}) {\left.
\right|_{r=a}}\equiv \mbox{\boldmath $\tau$} \cdot \mbox{\boldmath $
J$}^{\lambda  \, \beta}{\left. \right|_{r=a}} =  \,\nonumber \\
\left. \frac{1}{2}  \left[ \left( {\Psi}_\lambda ,
 \mbox{\boldmath $\tau$}  \cdot \mbox{\boldmath $\hat  V$}
{\Psi}_\eta \right) + \left( \mbox{\boldmath $\tau$}\cdot \mbox{\boldmath $\hat  V$}
{\Psi}_\lambda  , {\Psi}_\eta \right) \right] \right|_{r=a} = 0 \, ,
\label{generalflux}
\end{eqnarray}
where $a$ is the nanocrystal radius and  $\mbox{\boldmath
$\tau$}=\mbox{\boldmath $ r$}/r$ (in spherical coordinates
$\mbox{\boldmath $r$}$ is defined as $\mbox{\boldmath
$r$}\equiv(r,\Theta,\varphi )$).   This general requirement of Eq.
(\ref{generalflux}) should be satisfied for two arbitrarily chosen
eigenfunctions ${ \Psi}_\lambda $ and ${ \Psi}_\eta$   of the
Hamiltonian $\hat H^0$ defined in Eq. (\ref{h0})  with energies
$E_\lambda$ and $E_\eta$, respectively.  Substituting the explicit
expressions for the normal components of the envelope  velocities $
\mbox{\boldmath $\tau$}  \cdot \mbox{\boldmath $\hat
V$}=\mbox{\boldmath $\tau$}  \cdot \partial \hat H^0/\partial
\mbox{\boldmath $p$} $ into Eq. (\ref{generalflux}) one  obtains:
\begin{eqnarray}
&&J_{\tau}^{\lambda  \, \eta} (\mbox{\boldmath $r$})
\left|_{r=a}\right.  =
 \frac{i  \hbar}{2m_0} \left[ \alpha \left(\Psi^c_\eta
\mbox{\boldmath $\tau$} \cdot \mbox{\boldmath $\nabla$} \Psi^{c
*}_{\lambda } - \Psi^{c *}_{\lambda } \mbox{\boldmath $\tau$} \cdot
\mbox{\boldmath $\nabla$} \Psi^c_\eta \right)\right.\nonumber \\&&  +
\left. \left. \frac{2m_0 P}{\hbar}\left(  \mbox{\boldmath
$\tau$}\cdot {\bf \Psi}^{v}_\eta    \Psi^{c *}_{\lambda }  -
\mbox{\boldmath $\tau$}\cdot {\bf \Psi}^{v *}_{\lambda}
\Psi^{c}_\eta  \right) \right] \right|_{r=a} =0 \, . \label{fluxelec}
\end{eqnarray}
Using equation $\hat H^0 \Psi = E \Psi$  one can express the
components  of the spin vector ${\bf \Psi}^v$ through
$\mbox{\boldmath $\nabla$} \Psi^c$:
\begin{eqnarray}
{\bf \Psi}^v &= &\frac{\hbar}{2m_0P} \left( \alpha - \frac{m_0}{m_c(E)}
 \right)  \mbox{\boldmath $\nabla$} \Psi^c \nonumber \\
&+& \frac{i \hbar}{4 m_0P} \left( g_c(E) - g^* \right)
 [\mbox{\boldmath $\hat \sigma$} \times \mbox{\boldmath $\nabla$ }\Psi^c ]\, ,
\label{psiv}
\end{eqnarray}
where the energy dependent electron effective mass $m_c(E)$ and
$g$--factor  $g_c(E)$ are defined in Eqs. (\ref{bulkm}) and
(\ref{bulkg}) respectively. The normal projection $(\mbox{\boldmath
$\tau$} {\bf \Psi}^v)$ can  then be written:
\begin{eqnarray}
 \mbox{\boldmath $\tau$} {\bf \Psi}^v({\bf r}) &=&
\frac{\hbar}{2m_0P} \left( \alpha - \frac{m_0}{m_c(E)}  \right)
\frac{\partial \Psi^c}{\partial r}\nonumber \\
&-& \frac{ \hbar}{4 m_0 P r} \Delta g(E) (\mbox{\boldmath $\hat \sigma$}
 \mbox{\boldmath $\hat  L$}) \Psi^c \, ,
\label{psivtau}
\end{eqnarray}
where $\Delta g(E) = g^* - g_c(E)$. Substituting Eq. (\ref{psivtau})
into Eq. (\ref{fluxelec}) one obtains the  general requirement for
the conduction band component of the envelope wave function at the
NC surface:
\begin{widetext}
\begin{eqnarray}
 J_{\tau}^{\lambda  \, \eta}(a) =  \frac{i  \hbar}{2m_0}
\left[ \left(\frac{m_0}{m(E_\lambda)}
\Psi^c_\eta \frac{\partial \Psi^{c *}_{\lambda}}{\partial r} -
\frac{m_0}{m(E_\eta)} \Psi^{c *}_{\lambda } \frac{\partial
 \Psi^c_\eta}{\partial r} \right) + \right. \nonumber \\
\left. \left. \frac{1}{2r}
\left( \Psi^{c}_\eta   \Delta g(E_\lambda)(\mbox{\boldmath $\hat \sigma$}
 \mbox{\boldmath $\hat L$} ) \Psi^{c *}_{\lambda } - \Psi^{c *}_{\lambda }
  \Delta g(E_\eta)( \mbox{\boldmath $\hat \sigma$} \mbox{\boldmath $ \hat  L$})
   \Psi^c_\eta \right) \right] \right|_{r=a} =0. \label{fluxelecso}
\end{eqnarray}
\end{widetext}
Equation (\ref{fluxelecso}) must be fulfilled at each point of the
surface. It is satisfied for all arbitrary chosen energy states
$\lambda$ and $\eta$ if and only if the boundary conditions for
$\Psi^{c}_\lambda$ as well as for $\Psi^{c}_\eta$  are given by:
\begin{eqnarray}
\Psi^c(r=a,\Theta ,\varphi)  = Ta_0 \left( \frac{m_0}{m_c(E)}
\frac{\partial \Psi^c(r,\Theta ,\varphi)}{\partial r} + \right.
\nonumber \\ \left. \left.  \frac{1}{2r}\Delta g(E) (\mbox{\boldmath
$\hat \sigma \hat  L$})\Psi^c(r,\Theta ,\varphi) \right)
\right|_{r=a}  \, , \label{sphBC}
\end{eqnarray}
 where $T$ is a real number constant independent of $E$ and $a_0$ is
the lattice constant. We assume that the surface of the nanocrystal
also possesses the spherical symmetry and is characterized by the
same BCs at any point of the surface. In this case parameter $T$  in
the  GBCs of Eq. (\ref{sphBC}) is independent of angles $\Theta$ and
$\varphi $.

\subsection{\label{spherso}Electron energy levels:
 the  surface induced spin--orbit splitting}

Let us consider the effect of the GBCs of Eq. (\ref{sphBC}) on the
electron QSLs in the absence of a magnetic field. Using Eq.
(\ref{psiv}) that expresses the valence band component  ${\bf
\Psi}^v$ of the wave function through its conduction band component
$\Psi^c$ one can derive the bulk Schr\"{o}dinger equation describing
$\Psi^c$ inside a semiconductor nanocrystal:
\begin{equation}
- \,  \frac{\hbar^2}{2m_c( E)}\hat U_2 \mbox{\boldmath $\nabla$}^2
\Psi^c(\mbox{\boldmath $ r$}) =
 E \Psi^c(\mbox{\boldmath $ r$}) \, .
\label{Shr}
\end{equation}
One can see that each spin component of $\Psi^c$ is a solution of the
standard bulk Schr\"{o}dinger equation with the energy dependent
effective mass $m_c( E)$. Equation (\ref{Shr}) does not contain a
spin-orbit term and its solution can always be presented as
$f_l(r)Y_{l,l_z}(\Theta,\varphi)$, where $l$ and $l_z$ are the
angular momentum and angular momentum projection, correspondingly,
$Y_{l,l_z}(\Theta,\varphi)$ are the spherical harmonics, and the
radial function $f_l(r)$ satisfies to the following equation:
\begin{eqnarray}
- \,  \frac{\hbar^2}{2m_c( E)} \Delta_{l} f_l(r) =
 E f_l(r) \, , \label{Kaneel} \\   \Delta_{l}=
{\partial^2\over \partial^2 r}+{2\over r}{\partial\over
\partial r}-{{l}({l}+1)\over r^2}~ \, .
\nonumber
\end{eqnarray}

At the same time the GBCs of Eq. (\ref{sphBC}) mixes two electron
spin sublevels, and all electron states are characterized now by the
total angular momentum: $\mbox{\boldmath $ j$}=\mbox{\boldmath $ L$}
+ \mbox{\boldmath $S$}$.  As a result  $\Psi^c$ is an eigen spinor of
the operator of the total angular  momentum:
\begin{eqnarray}
\Psi^c=f_l^\pm(r) \Omega_{j,l,m}(\Theta,\varphi ) \, ,
\label{psic}
\end{eqnarray}
where $\Omega_{j,l,m}(\Theta,\varphi )$ are the spherical spinors,
$j$ and $m$  are the  total angular momentum  and its projection
respectively, and $l$ is the angular momentum of the electron state.
The two radial functions $f_l^{\pm}$ describe the two electron states
with $j=l\pm 1/2$  (note that the states described by $f_l^-$ exist
only for $l \ge 1$). We use the definition of the spherical spinors
as in Ref. \onlinecite{ll} \begin{eqnarray} \Omega_{j,l,m}=\left(
\begin{array}{c}
C_{1/2,1/2;l,m-1/2}^{j,m} Y_{l,m-1/2} \\
C_{1/2,-1/2;l,m+1/2}^{j,m} Y_{l,m+1/2} \end{array} \right) \, ,
\end{eqnarray}
where $j=l\pm 1/2$ and $C_{j1,m1;j2,m2}^{j,m}$ are the Clebsch-Gordan
coefficients. Substituting Eq. (\ref{psic}) into Eq. (\ref{sphBC})
one can write  the general boundary condition for the radial  wave
functions $f_l^\pm(r)$:
\begin{eqnarray}
f_l^{\pm}(a)= T a_0 \left[ \frac{m_0}{m_c(E)} f_l^{\pm \, '}(a) +
\delta k_l^\pm  f_l^{\pm}(a) \right] \, , \label{bcfso}
\end{eqnarray}
where
\begin{eqnarray}
\delta k_l^+ = \frac{l}{2a}\Delta g(E) \,  , \quad l=0,1,2 \dots \, ,
\end{eqnarray}
 for the states with $j=l+1/2$ and
\begin{eqnarray}
\delta k_l^- = -\frac{l+1}{2a}\Delta g(E)\, , \quad l=1,2,3 \dots \, ,
\end{eqnarray}
for the states with $j=l-1/2$.
The conventionally used standard BCs for the impenetrable barrier
assumes vanishing of the wave function $\Psi^c$ at the nanocrystal
surface,  and thus correspond to $Ta_0=0$. In general, however,
the conduction band component of the wave function  does not vanish
at the surface.  Furthermore, it has been shown in
Ref.\onlinecite{bc}  that the boundary  condition with $Ta_0=0$ is
incorrect in cases where interband coupling is important. The
appropriate value of $Ta_0$ should  be determined from the fitting
to the experimental data.

The solutions of  Eq. (\ref{Kaneel})  can be written as  $f_l^\pm(r)
= (C_l^\pm/a^{3/2}) j_l(\phi_{l,n}^\pm r/a)$, where $j_l$ are
spherical Bessel functions, $C_l^\pm$ is the normalization constant
determined by the condition $\int (|\Psi^c|^2 + |{\bf \Psi}^v|^2)d{
V} = 1$, and  $\phi_{l,n}^\pm$ is connected with the energy, $E$, of
the electron level: $E=\hbar^2 \phi_{l,n}^{\pm 2}/2m_c(E)a^2$. An
equation that determined $\phi_{l,n}^\pm$ is obtained by substitution
of  $f_l^\pm(r)$ into the BCs of Eq. (\ref{bcfso}):
\begin{eqnarray}
 &&j_{l}(\phi_{l,n}^\pm) \left[1 - \delta k_l^\pm Ta_0 \right] = \label{bcl}  \\
\frac{T a_0}{a} \frac{m_0}{m_c(E)}
 \phi_{l,n}^\pm &&\left[
\frac{l}{2l+1} j_{l-1}(\phi_{l,n}) - \frac{l+1}{2l+1} j_{l+1}(\phi_{l,n})
\right] \,  \, . \nonumber
\end{eqnarray}
The $n^{\rm th}$ solution of this equation defines the energy of the
$n^{\rm th}$ electron level with the angular momentum $l$ and the
total angular momentum $j=l\pm 1/2$. In the case of the standard BCs
$Ta_0=0$, and the solution $\phi_{l,n}^\pm$ is the $n$--th zero
$\phi_{l,n}^0$ of the spherical Bessel functions $j_{l}$. The general
BCs with $Ta_0 \ne 0$ takes into account the effect of the surface on
the electron energy levels that is important in small nanocrystals.
In large nanocrystals satisfying the condition $a \gg |Ta_0|
(m_0/m_c)$, the solutions $\phi_{l,n}^\pm$ are close to
$\phi_{l,n}^0$ and the effect of the surface on the confined electron
states is negligible.

The electron states with $j=l+1/2$ and $j=l-1/2$ have different
energies for  $l \ge 1$ as a result of the general BCs of Eq.
(\ref{bcfso}).  This difference describes the surface induced
spin--orbit splitting of the excited states: $\Delta_c= E_{l+1/2} -
E_{l-1/2}$. It can be shown that $\Delta_c$ is positive if the
spin--orbit splitting of the valence band $\Delta>0$.

\begin{figure*}[hp]
\begin{center}
\includegraphics*[width=8cm,height=11cm]{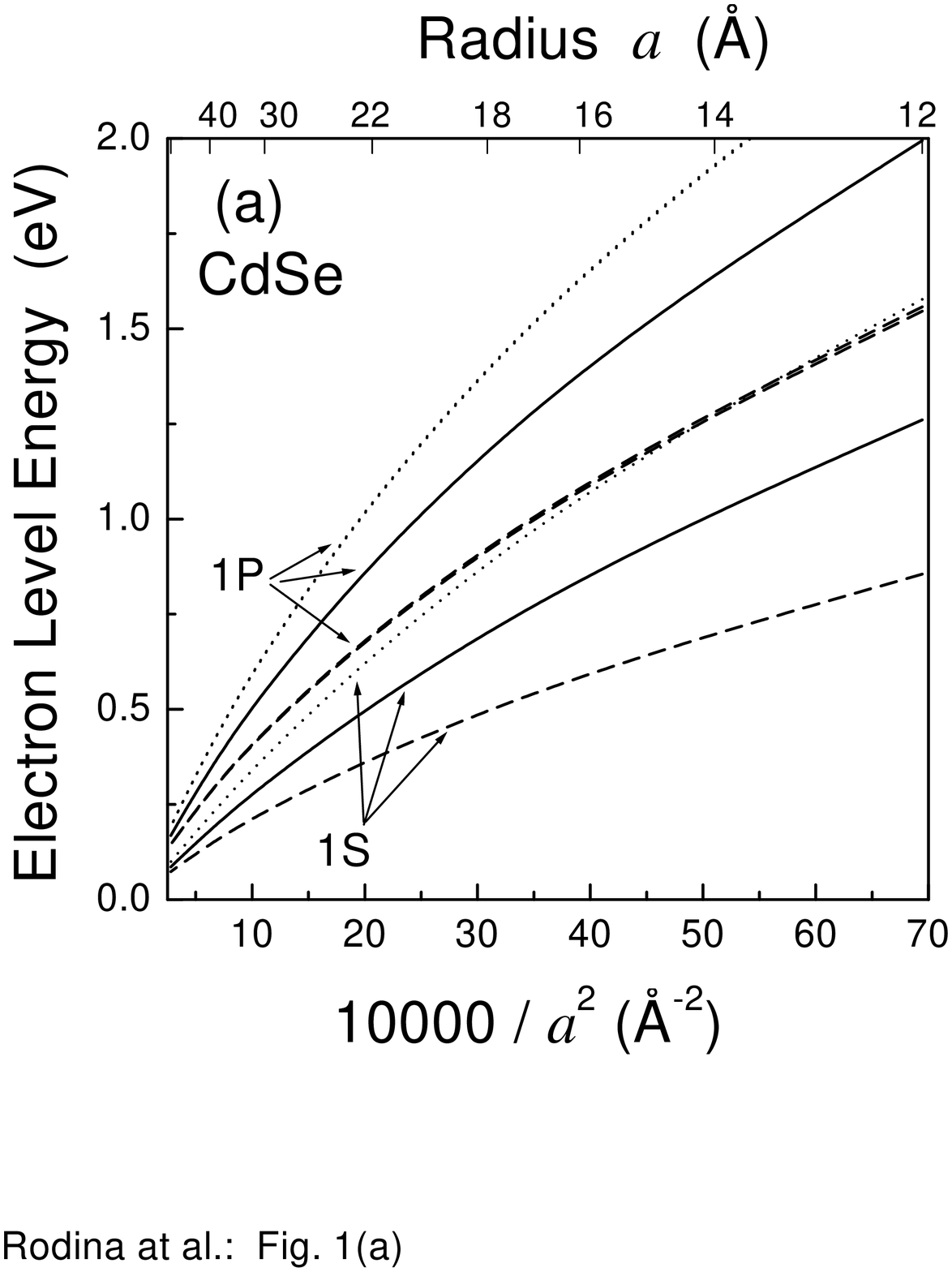}
\includegraphics*[width=8cm,height=11cm]{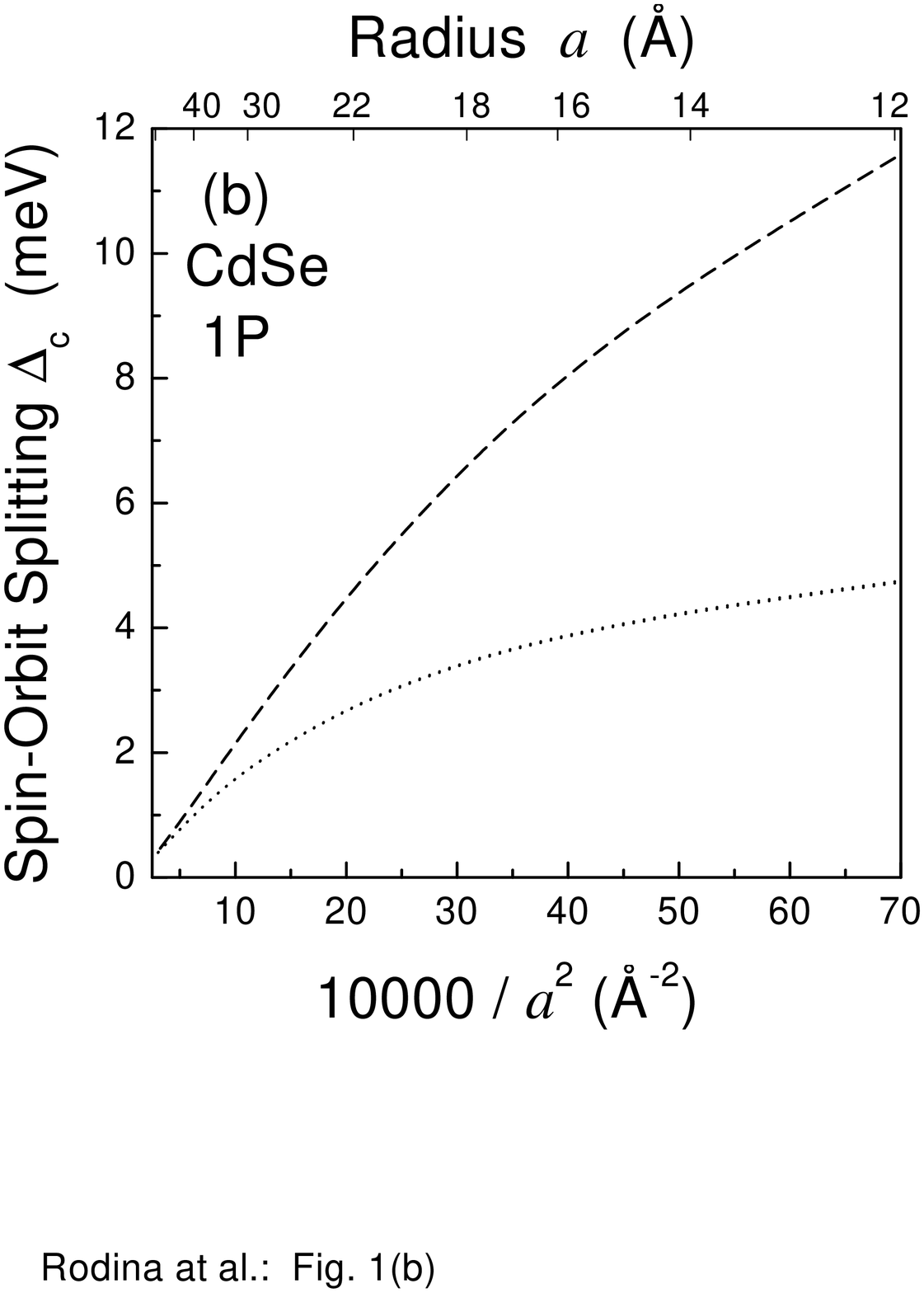}
\end{center}
\caption{\label{levels} The effect of general boundary conditions on
the ground $1S$ and first excited $1P$  energy levels  (a), and on
the spin--orbit splitting between $1P_{3/2}$ and $1P_{1/2}$ levels
$\Delta_c$   (b) in bare CdSe nanocrystals. The size dependencies are
calculated for the standard BCs with the surface parameter $Ta_0= 0$
$\AA$ (solid line), for  $Ta_0=-0.6$ $\AA$ (dashed lines), and for
$Ta_0=0.6$ $\AA$ (dotted lines). The bulk parameters of CdSe used in
calculations are described in the text.}\end{figure*}

Figure \ref{levels} shows the effect of the general BCs on the energy
 of the ground $1S$ ($l=0$) and the first excited $1P$ ($l=1$)
electron states in bare CdSe nanocrystals. The surface induced
spin-orbit interaction splits the last state into two states with
the total angular momentum $j=3/2$ ($P_{3/2}$ state) and $j=1/2$ ($P_{1/2}$
state) and the size dependence of this splitting
$\Delta_c=E(1P_{3/2})-E(1P_{1/2})$ is shown in Fig. \ref{levels}(b).
The calculations have been performed for the  following bulk
parameters of CdSe: $E_p=19.0$ eV and  $m_c(0)=0.116$ $m_0$ from
Ref. \onlinecite{seisyan}, $E_g=1.839$ eV, $\Delta=0.42$ meV, and
$g_c(0)=0.68$ from Ref. \onlinecite{Landoldt}, that result in
$\alpha=-1.07$ and $g^*=1.96$. One can see that even a small surface
parameter introduced by the GBCs, $|Ta_0|=0.6$ $\AA$,  significantly
affects both ground and excited states in small nanocrystals. The
positive (negative) value of the surface parameter $Ta_0 >0$
($Ta_0<0$) increases (decreases) the energy of all states in
comparison with $Ta_0=0$.   The relative order of the
$P_{3/2}$ and  $P_{1/2}$ states with the total momentum $j=3/2$ and
$j=1/2$  always coincides with the relative order of the
valence band subbands characterized by the same total momentum. The
value of the spin-orbit splitting, $\Delta_c$, however, remains for
all excited states  much smaller than the averaged energy
$E_l=(lE_{l-1/2}+(l+1)E_{l+1/2})/(2l+1)$ of the level  with the
orbital momentum $l$.

If $|\delta k_l^\pm Ta_0| \ll 1$, the averaged energies $E_l$ and
corresponding wave numbers $\phi_{l,n}$ can be found from the
simplified BCs given by Eqs. (\ref{bcfso}, \ref{bcl}) with $\delta
k_l^\pm = 0$. The  energy corrections $\Delta E_l^\pm$ coming from
the small $\delta k_l^\pm \ne 0$  can be found perturbatively.  In
the Appendix we obtain the following energy corrections:
\begin{eqnarray}
\Delta E_{l}^\pm =  \frac{\hbar^2}{2m_0} |f_l^\pm(a)|^2 a^2 \delta
k_l^\pm = \nonumber \\ \frac{\hbar^2}{4m_0} \int  \left( \Delta g(E)
\Psi^{c *}( \mbox{\boldmath $ \hat \sigma \hat L$})  \Psi^c \right)
\frac{1}{r} dS \, , \label{sosur}
\end{eqnarray}
where $dS =  r^2 sin(\Theta )d\Theta d\varphi$. Note that although
$\phi_{l,n}^+=\phi_{l,n}^-=\phi_{l,n}$, the normalization constants
$C_l^+ \ne C_l^-$ . However, in the considered approximation they
can be replaced with high accuracy by the averaged constant
$C_l=C_l^{+}=C_l^{-}$ that is determined by the approximate
normalization condition:
\begin{eqnarray}
&&\int_0^a |f_l|^2 r^2dr + \int_0^a|\Phi_l|^2 r^2dr = 1 \, , \\
|\Phi_l|^2&=& \frac{\hbar^2}{2m_0E_p}
\left(\alpha-\frac{m_0}{m_c(E)}\right)^2\left(|f_l^{'}|^2  +
\frac{l(l+1)}{r^2}|f_l|^2 \right) \, , \nonumber
\end{eqnarray}
 that neglects the second term in Eq. (\ref{psiv}). Thus, if $|\delta
k_l^\pm Ta_0| \ll 1$,  $\Delta_c < E_l$, the energy levels of the
excited states $E_l^{\pm}$ with $j = l \pm 1/2$ are the eigen
energies of the effective Hamiltonian $\hat H_l$:
\begin{eqnarray}
\hat H_{l} \Omega_{l \pm 1/2,l,m} = E_l^\pm \Omega_{l \pm 1/2,l,m}  \, , \nonumber \\
\hat H_{l} = E_l\hat U_2 +  \frac{1}{2l+1} \Delta_c (E_l)  \left( \mbox{\boldmath $
\hat L \hat  \sigma$} \right)
\label{hlso}
\end{eqnarray}
where the  energy dependent spin--orbit splitting $\Delta_c(E_l)$ is given by
\begin{eqnarray}
\Delta_c(E_l) = \left(l + \frac{1}{2} \right) \frac{\hbar^2}{2m_0a^2}
\Sigma_{sur}(E_l) \, ,
 \label{deltac} \\ \Sigma_{sur}(E_l)=\Delta g(E_l) a^3 {|f_l(a)|}^2=
\Delta g(E_l) C_l^2 j_{l}^2(\phi_{l,n}) \, . \label{sigmasur}
\end{eqnarray}
The approximate expressions of  Eqs. (\ref{deltac},\ref{sigmasur})
describe the exact splitting $\Delta_c(E_1)$ of the  $1P$ states in
CdSe nanocrystals  shown in Fig. \ref{levels}(b) with high  accuracy
(the largest error in the smallest nanocrystals is about $2 \%$  for
$Ta_0 = - 0.6$ $\AA$). It is important to note here that while the
matrix elements of the spin--orbit operator $( \mbox{\boldmath $ \hat
L \hat \sigma$})$ are nonzero only for $l \ge 1$,   Eq.
(\ref{sigmasur}) defines also the surface parameter $\Sigma_{sur}$
for the  $l=0$ $S$ symmetry states.

The dimensionless parameter $\Sigma_{sur}$ is proportional to the
spin--orbit splitting in the valence band $\Delta$ and to the square
of the wave function at the surface. It describes the
surface--induced spin--orbit coupling of the conduction electron QSLs
in bare semiconductor nanocrystals. In bulk semiconductors the
integration in Eq. (\ref{sosur}) results in zero spin--orbit
splitting of the conduction band states because it must be carried
out at the remote bounding surface where the wave function is
vanishing. $\Sigma_{sur}=0$  if one assumes the vanishing of the
conduction band component wave function at the surface of spherical
nanocrystals. This assumption is never justified if the coupling
between conduction band and valence band components is significant.
Thus, the spin--orbit splitting of the electron QSLs is caused by the
admixture of the valence band states near the surface.

\section{\label{magnetic}The fine structure of electron QSLs in a magnetic field}

\subsection{\label{msur}Surface induced magnetic moment of the confined electrons}

Let us consider now the energy of the electron QSLs, $E_l^\pm$,  in a
weak magnetic field.  Substituting  the zero field wave functions
$\Psi= \{\Psi^c, {\bf \Psi}^v \}$ given by Eq. (\ref{psic}) and
(\ref{psiv}) into  Eqs. (\ref{dec}) and (\ref{dcoup}), we obtain for
the energy correction $\Delta E \approx \Delta E_c + \Delta
E_{coup}$:
\begin{eqnarray}
\Delta E = \mu_B H \left[ \frac{1}{2} \left< {\Psi}^c | g_c(E)
(\mbox{\boldmath $\hat \sigma n$}) |  { \Psi}^c \right> + \right.
\nonumber \\ \left.
 \left< {\Psi}^c |\frac{m_0}{m_c(E)}( \mbox{\boldmath $ \hat  L n$})
|  { \Psi}^c \right> \right] + \Delta E_{sur}\, , \label{split}
\end{eqnarray}
where we neglect $\Delta E_v$ corrections of Eq. (\ref{dev}) as being
smaller by the factor $E_l/(E_l+E_g)$. The second term of Eq.
(\ref{split}) describes the unexpected surface contribution to the
electron magnetic energy:
\begin{eqnarray}
\Delta E_{sur} =  \frac{\mu_B H}{4}   \int dS \frac{1}{r} \Delta
g(E_l) (\Psi^{c *}  [ \mbox{\boldmath $ r$} \times [\mbox{\boldmath
$\hat \sigma$} \times \mbox{\boldmath $ r$}]] \Psi^c) \,.
\end{eqnarray}
The  effective Hamiltonian describing the fine structure of the
electron  state with the orbital momentum $l$ in a magnetic field can
be written as $\hat H_{l} + \hat H_{H}$, where $\hat H_{l}$ is
defined by Eq. (\ref{hlso}) and the effect of a weak magnetic field
is described by:
\begin{eqnarray}
\hat H_{H}=  \frac{1}{2}\mu_B \bar g_s(E_l) (\mbox{\boldmath $ \hat
\sigma  H$})  + \mu_B \bar g_l(E_l) (\mbox{\boldmath $ \hat  L H $})
+ \nonumber \\ \frac{1}{4}\mu_B
\Sigma_{sur}(E_l)\left([\mbox{\boldmath $\tau$} \times
[\mbox{\boldmath $\hat \sigma$} \times \mbox{\boldmath $\tau$}]
\mbox{\boldmath $ H$} \right)  \, . \label{zeeman}
\end{eqnarray}
Here the weighted spin $\bar g_s$ and orbital $\bar g_l$ $g$ factors
are given by
\begin{eqnarray}
 \bar g_s &=& \int_0^a g_c(E_l)  r^2 dr |f_l(r)|^2  \, , \label{bargs} \\
  \bar g_l &=&\int_0^a m_0/m_c(E_l) r^2 dr |f_l(r)|^2 \, ,
\end{eqnarray}
 respectively.  The first two terms in Eq. (\ref{zeeman}) describe
 the averaged volume energy of the  spin $\mbox{\boldmath $\mu$}_S= -
 \mu_B \bar g_s(E_l) \mbox{\boldmath $ S$}$ and  orbital
 $\mbox{\boldmath $\mu$}_L=-\mu_B \bar g_l(E_l)\mbox{\boldmath $\hat
 L$}$ magnetic moments in an external magnetic field, and the last
 term describes the energy of the surface--induced magnetic moment
 $\mbox{\boldmath $\mu$}_{sur}(E_l)$   given by:
\begin{eqnarray}
\mbox{\boldmath $\mu$}_{sur}= - \frac{\mu_B}{4} \Sigma_{sur}(E_l)
[\mbox{\boldmath $\tau$} \times [\mbox{\boldmath $\hat \sigma$}
\times \mbox{\boldmath $\tau$}]  \, .
 \end{eqnarray}
This magnetic moment  arises from  the surface--induced spin--orbital
term of Eq. (\ref{hlso}) which is modified  by the Larmor precession
of  the electron in an external magnetic field. Indeed, the surface
energy term $\Delta E_{sur} = -(\mbox{\boldmath $\mu$}_{sur}
\mbox{\boldmath $ H$})$ can be obtained directly by replacing  $-i
\mbox{\boldmath $\nabla$ } \rightarrow -i \mbox{\boldmath $
\nabla$}+(e/c\hbar) \mbox{\boldmath $ A$}$ in the spin--orbit
perturbation term $(\mbox{\boldmath $\hat  L \hat \sigma$})= -i
([\mbox{\boldmath $ r$} \times \mbox{\boldmath $\nabla$}]
\mbox{\boldmath $\hat \sigma$} )$ of Eq. (\ref{hlso}).  The origin of
the surface--induced electron magnetic moment is similar to those of
an additional relativistic magnetic moment of the electron in atoms,
and their values  are derived in a similar way (see for example Refs.
\onlinecite{bethe,abragam,margenau}).
 It is necessary to note that the surface-induced magnetic moment
 $\mbox{\boldmath $\mu$}_{sur}$ does not vanish for the states with
$l=0$ even when it arises from the spin--orbit coupling term
 $(\mbox{\boldmath $\hat L \hat \sigma$})$.

\subsection{\label{ground}Size dependence of the ground state electron $g$ factor}

\begin{figure*}[hp]
\begin{center}
\includegraphics*[width=8cm,height=11cm]{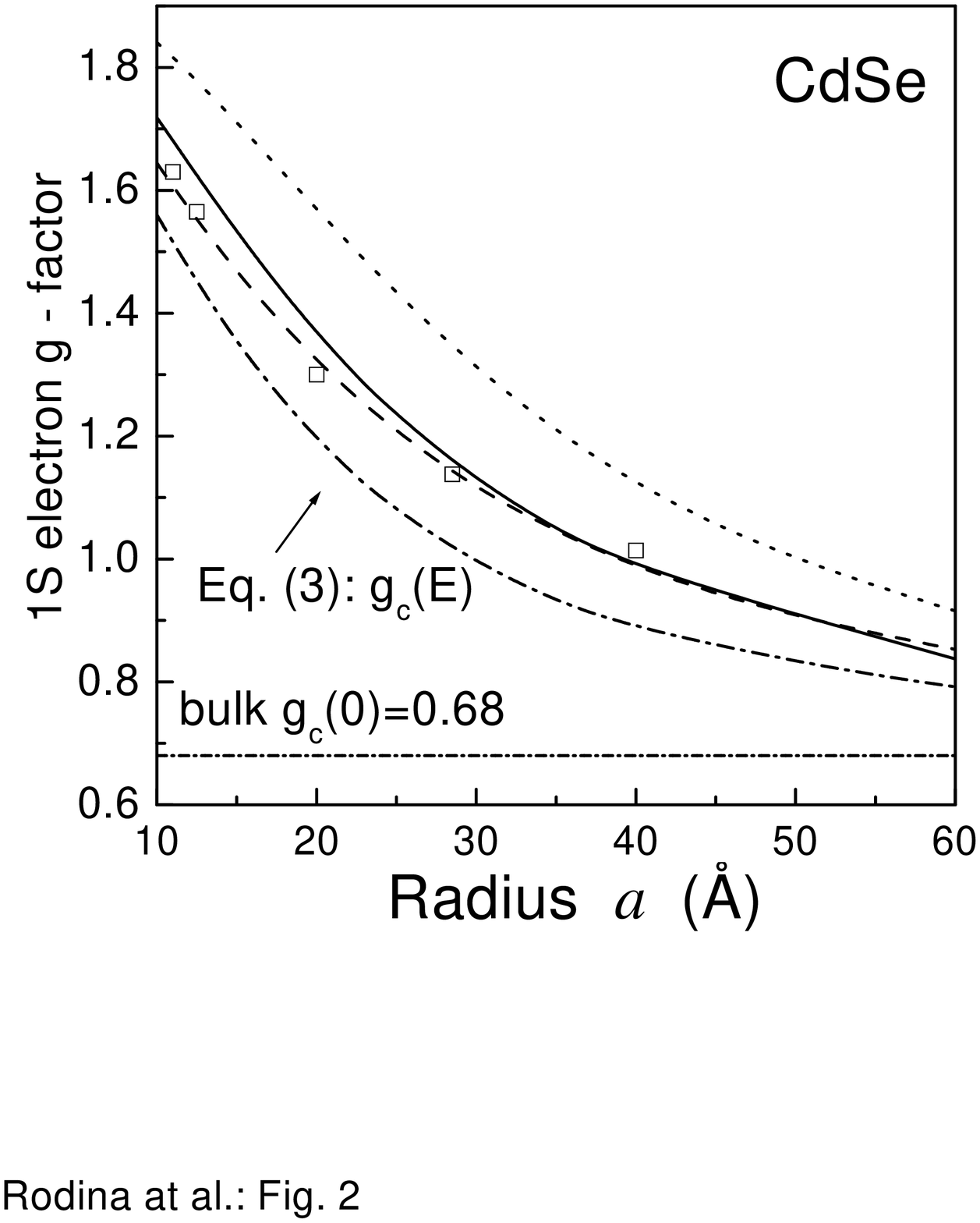}
\end{center}
\caption{\label{sizegfact}The size dependence of the  ground $1S$
state electron  $g$-- factor in bare CdSe nanocrystals  calculated
for the standard BCs with the surface parameter $Ta_0=0$ (solid
line), for $Ta_0=-0.6$ $\AA$ (dashed line), and  for $Ta_0=0.6$ $\AA$
(dotted line). The empty squares show the experimental data from Ref.
{\protect \onlinecite{gupta02}}. For comparison we also show the
energy dependence of the bulk electron g-factor $g_c(E)$ calculated
at the energy of the $1S$ electron level (dash--dotted line) and bulk
value of the electron $g$-factor (short dash--dotted line).}
\end{figure*}

All electron states that have $S$ symmetry are Kramers doublets  that
are degenerate with respect to its spin projection. The external
magnetic field lifts this degeneracy and splits these states into two
levels with energy:
\begin{equation}
E_0(H) = E_0 \pm \frac{1}{2} \mu_B g_s(E_0) H \, ,
\end{equation}
where the $\pm$  signs correspond to the electron states with spins
parallel   and antiparallel to  the magnetic field direction. The
spin effective $g$ factor can be obtained from Eq. (\ref{zeeman}) as:
\begin{eqnarray}
g_s(E_0) =  \bar g_s(E_0) + \frac{1}{3} \Sigma_{sur} \, .
 \label{gs}
\end{eqnarray}
The effective electron $g$ factor for the $S$ electron states in
spherical heterostructures, $g_s(E_0)$, was first obtained in Ref.
\onlinecite{kiselev} where  the standard BCs at the heterointerface
were used to describe the size dependence of the electron $g$
factor.  In the case of the bare semiconductor nanocrystals  the
standard BC at the surface ($f_0(a)=0$)   would lead to  zero
surface contribution to the effective electron $g$ factor.  We have
examined here the effect of the general BC of Eq. (\ref{bcfso}) on
the electron effective $g$ factor in CdSe nanocrystals.  The
weighted  $g$ factor of the states with $S$ symmetry in
Eq. (\ref{gs}) can be written:
\begin{equation}
\bar g_s(E) = g_0 + \int_0^a r^2 dr |f_l(r)|^2 \left(g_c(E)-g_0\right) \, ,
\end{equation}
where the corrections $\Delta E_v$ of Eq. (\ref{dev}) to the $\bar
g_s$ of Eq. (\ref{bargs}) have been added. Figure \ref{sizegfact}
shows the size dependence of the ground state electron $g$ factor,
$g_s(E_0)$, calculated using the standard BC with $|Ta_0|=0$ and the
general BC with $|Ta_0|=0.6$ $\AA$. One can see that the
experimental  size dependence of the electron $g$-factor taken from
Ref. \onlinecite{gupta02}  is described very well by the general BC
with the negative parameter $Ta_0=-0.6$,  while the use of the positive
parameter brings the theoretical curve far away from the data.

\begin{figure*}[hp]
\begin{center}
\includegraphics*[width=8cm,height=11cm]{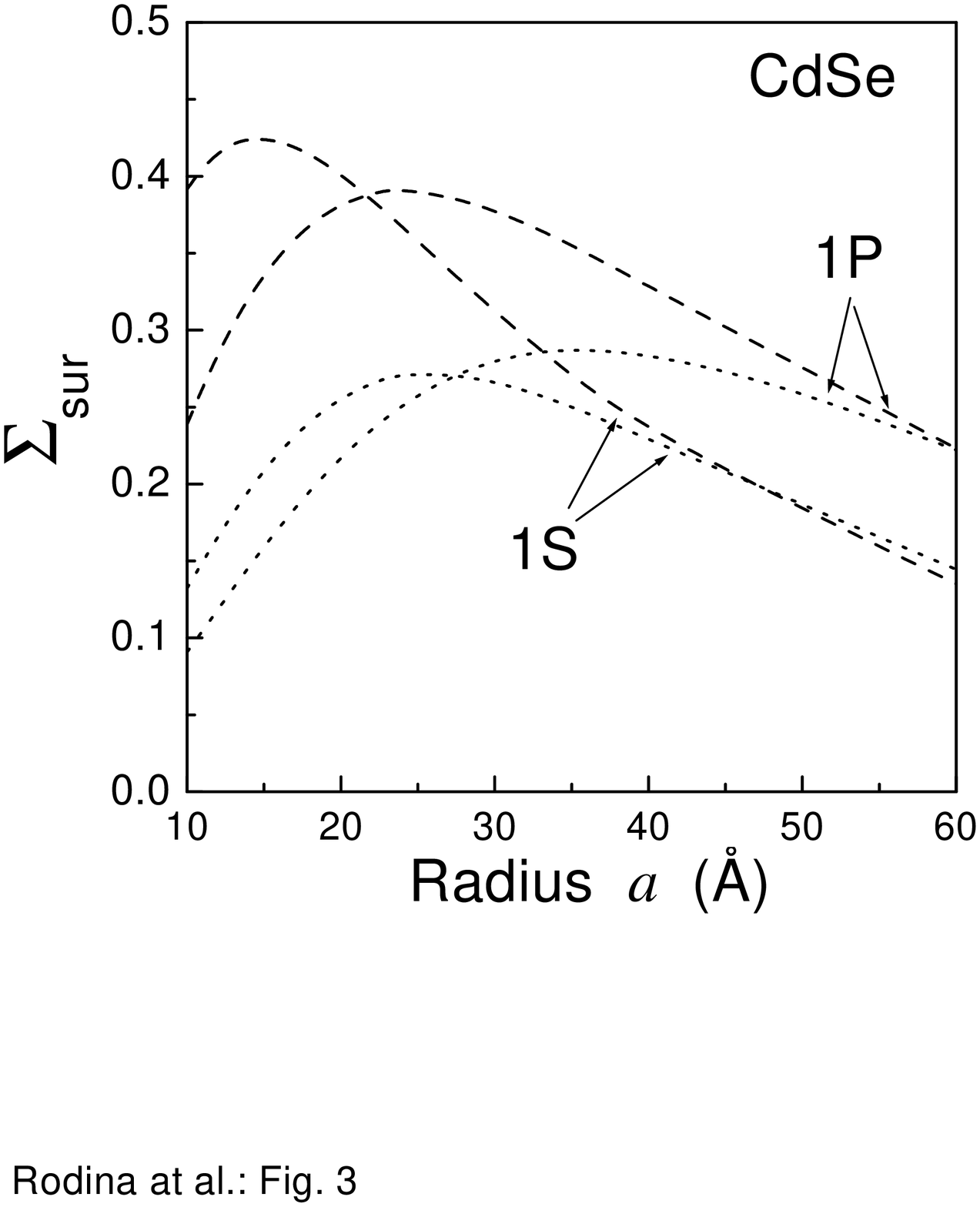}
\end{center}
\caption{\label{sizesurf}The size dependence of the dimensionless
surface parameter $\Sigma_{sur}$ of Eq. (\ref{sigmasur}) for the
ground $1S$ ($l=0$) and first excited $1P$ ($l=1$) states in bare
CdSe nanocrystals. The dashed  and dotted curves are calculated with
the surface parameters $Ta_0=-0.6$ $\AA$  and  $Ta_0=0.6$ $\AA$
respectively.}
\end{figure*}

The size dependence of the bulk--like energy--dependent $g$ factor,
$g_c(E_0)$, calculated with the help of Eq. (\ref{bulkg}) for
$Ta_0=0$ is also shown in Fig. \ref{sizegfact}. One can see that
the bulk  $g$--factor $g_c$ fails to describe the experimental data
even in the largest dots. At the same time, the difference between two
$g_s$ curves calculated with  $Ta_0=-0.6$ $\AA$ and $Ta_0= 0$ is not
very large. This is because the general BCs affects the electron
g-factor in two ways: indirectly, through the change of the zero
field energy $E_0$ decreasing/increasing the electron g-factor, and
directly, through the surface contribution $1/3 \Sigma_{sur}(1S)$,
that is always positive. In the case of the negative $Ta_0<0$  the
GBCs decreases the energy of the ground electron state decreasing the
electron g-factor, and these two effects partly compensate for each other.
However, the surface contribution  $1/3 \Sigma_{sur}(1S)$ calculated
with $Ta_0 = -0.6$ $\AA$ is significant in small nanocrystals (see
Fig. \ref{sizesurf}).

\subsection{\label{excited}Symmetry of the confined electron
 excited states in an external magnetic field}

The fine structure of the excited electron states in an external
magnetic field is very sensitive to  the relation between
$\Delta_c(E_l)$ and the energy of the orbital magnetic  moment $\bar
g_l \mu_B H$. In the {\it low--field regime}, when the magnetic
energy is smaller than the separation between the electron states
with $j=l \pm 1/2$, the external  magnetic field satisfying the
condition $\bar g_l \mu_B H \ll \Delta_c$ splits the levels according
to projection $m$ of the full momentum $j$ on the magnetic field as:
\begin{eqnarray}
E_{j,m}^{+} = E_l + \Delta_c(E_l) \frac{l}{2l+1} +
 \mu_B \, g_{j}^{+}(E_l)  \, m \, H  \, ,
\end{eqnarray}
for  $j=l+1/2$ and
\begin{eqnarray}
E_{j,m}^{-} = E_l  - \Delta_c(E_l) \frac{l+1}{2l+1} +
 \mu_B \, g_{j}^{-}(E_l) \, m \, H \, .
\end{eqnarray}
for $j=l-1/2$.
 The effective electron $g$--factors
\begin{eqnarray}
g_{j}^+&=&\frac{1}{2j}\bar g_s ( E_l) + \frac{ 2j-1}{2j} \bar
g_l(E_l) + \frac{2j+1}{8j(j+1)}\Sigma_{sur}  \, ,
 \label{elgfe} \\
g_{j}^-&=&\frac{-1}{2(j+1)} \bar g_s ( E_l) +
\frac{ 2j + 3}{2(j+1)} \bar g_l(E_l) -\frac{2j+1}{8j(j+1)}\Sigma_{sur}
\nonumber \\&& \label{elgfo}
\end{eqnarray}
are the analogs of the Lande factors for electrons in atoms in the
case of  the "anomalous" Zeeman effect.\cite{bethe} The surface
contributions to the electron $g$--factors  ($\propto \Sigma_{sur}$)
are similar to the  relativistic corrections in Refs.
\onlinecite{abragam,margenau}.

When the zero field fine structure splitting becomes smaller than the
energy of the orbital magnetic moment, the magnetic field Zeeman term
mixes the states with different $j=l \pm 1/2$. In this case the
projections of the spin momentum $s_z$ and orbital momentum $l_z$ on
the direction of the magnetic field are a more convenient notation
for describing the electron state fine structure. The spin--orbit
term $(\mbox{\boldmath $ L S$})$ and the additional surface moment
$(\mbox{\boldmath $\mu$}_{sur} \mbox{\boldmath $ H$})$ may, however,
mix the states with different values of the $s_z l_z$ product.

\begin{figure*}[hp]
\begin{center}
\includegraphics*[width=8cm,height=11cm]{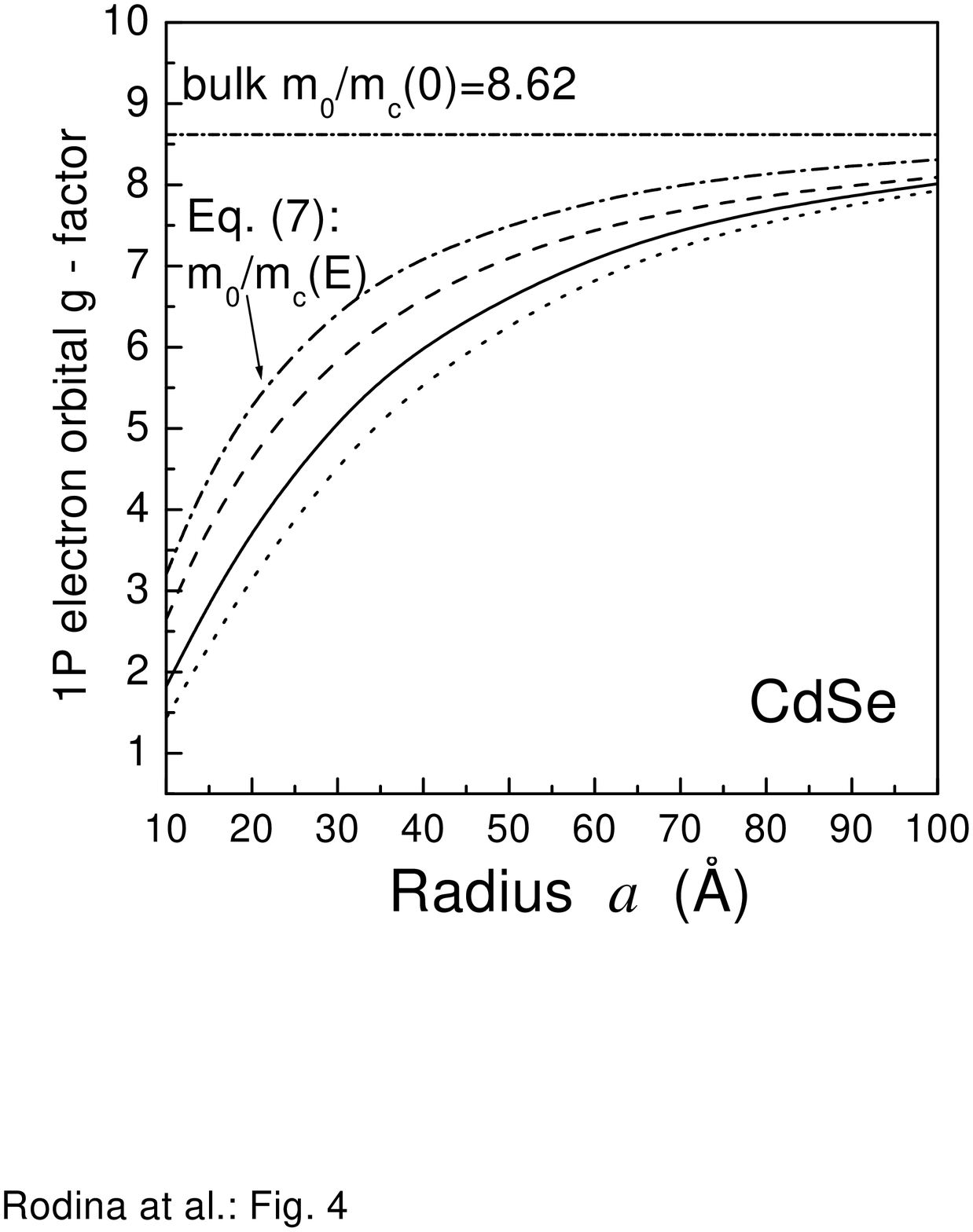}
\end{center}
\caption{\label{sizegl}The size dependence of the   orbital  $g$--
factor, $\bar g_l(1P)$, for the first excited state in bare CdSe
nanocrystals calculated for the standard BCs with surface parameter
$Ta_0=0$ (solid line), for  $Ta_0=-0.6$ $\AA$ (dashed lines), and for
$Ta_0=0.6$ $\AA$ (dotted lines). For comparison we also show the bulk
energy dependence of the bulk orbital $g$-- factor $m_0/m_c(E)$
calculated at the energy of the $1P$ electron level (dash--dotted
line) and its value at the bottom of the conduction band (short
dash--dotted line).}
\end{figure*}

Let us consider the matrix elements of the surface magnetic moment
operator $\mbox{\boldmath $\mu$}_{sur}$ on the eigen-functions of one
angular momentum $l$. Assuming that the states with different $l$ are
not mixed, one can find:
\begin{eqnarray}
\frac{r_i r_j}{r^2} Y_{l,m}(\Theta ,\varphi ) = \left(
\frac{2l-1+2l^2}{(2l+3)(2l-1)} \, \delta_{ij} -  \right. \nonumber \\
\left. \, \frac{2}{(2l+3)(2l-1)} \{L_i L_j\} \right) Y_{l,m}(\Theta
,\varphi ) \, , \label{rij}
\end{eqnarray}
where $i,j=x,y,z$ and $\delta_{ij}$ is the Kronecker delta symbol.
Substituting Eq. (\ref{rij}) into $( \mbox{\boldmath $ H$}
[\mbox{\boldmath $\tau$} \times [\mbox{\boldmath $\hat \sigma$}
\times \mbox{\boldmath $\tau$}] )= (\mbox{\boldmath $ H \hat
\sigma$}) - (\mbox{\boldmath $ H \tau$})(\mbox{\boldmath $\tau \hat
\sigma$})$, we rewrite the magnetic field Hamiltonian $\hat H_H$ of
Eq. (\ref{zeeman}) as:
\begin{eqnarray}
\hat H_H =  \mu_B g_s(E_l) (\mbox{\boldmath $ S H$})  + \mu_B \bar
g_l(E_l) (\mbox{\boldmath $   L H$}) + \nonumber \\ \mu_B
\Sigma_{sur}(E_l) \frac{(\mbox{\boldmath $ S H$}) \hat L^2 +
\{(\mbox{\boldmath $S L$})(\mbox{\boldmath $ L H$})\}}{(2l+3)(2l-1)}
\, . \label{zeemanpb}
\end{eqnarray}
Here the spin electron $g$ factor is given by
\begin{equation}
g_s(E_l) = \bar g_s(E_l)-  \frac{1}{(2l+3)(2l-1)}\Sigma_{sur}(E_l) \, .
\end{equation}
The matrix elements of the last term in Eq. (\ref{zeemanpb}) are zero
for $l=0$.  If $\Sigma_{sur}(E_l) \ll \bar g_l(E_l)$ one can neglect
the off diagonal elements of the operator $\{(\mbox{\boldmath $ S
L$})(\mbox{\boldmath $ L H$})\}$ and replace them with
$(\mbox{\boldmath $ S H$})(\mbox{\boldmath $ L n$})^2$. This is the
case for the first excited $1P$ state in CdSe nanocrystals (for
comparison, see the corresponding curves in Figs. \ref{sizesurf} and
\ref{sizegl}).  In this case the last  term  of Eq. (\ref{zeemanpb})
describes the surface contribution to the spin splitting of the
electron levels that depends on the projection of the orbital
momentum on the magnetic field.\cite{errow}

In the  {\it strong field regime} (similar to the case of the
"quasi--normal" Zeeman effect or "complete" Paschen-Back effect for
the electrons in atoms\cite{bethe}) one can additionally  neglect the
off diagonal elements of the spin--orbit operator $(\mbox{\boldmath $
L S$})$. The fine structure of the electron level with orbital
momentum $l$  is described by:
\begin{eqnarray}
&&E_{l,l_z,s_z}= E_l +\frac{2s_z l_z}{2l+1} \Delta_c (E_l) + \\ &&\mu_B\,
s_z g_s(E_l,l_z)\, H + \mu_B l_z \bar g_l(E_l) \,H \, , \nonumber
\end{eqnarray}
where $l_z=-l,...0,...l$ and $s_z=\pm 1/2$ are the projection of the
electron angular momentum and spin on the magnetic field direction
and
\begin{equation}
g_s(E_l,l_z) = \bar g_s(E_l)+  \frac{l^2+l+l_z^2-
1}{(2l+3)(2l-1)}\Sigma_{sur}(E_l) \, .
\end{equation}

Thus the structures of the excited QSLs with $l \ge 1$ may be
different in weak and strong magnetic fields and undertake the
anticrossing in the intermediate magnetic field (similar to those
known for atoms\cite{bethe}). In nanocrystals
 the weak, intermediate and strong field regimes depend strongly on
the nanocrystal size as well as on its surface conditions.

\section{\label{sum}Discussion and Conclusion}

We have studied the effect of general boundary conditions (the
surface effect) on the electron QSLs  and their Zeeman splitting in
spherical bare semiconductor nanocrystals.  The above consideration
that has been carried on in the eight--band effective mass model can
be easily  extended to describe spherically layered  heterostructures
and wide gap semiconductor nanocrystals, which are better described
by the fourteen--band model (see for example Ref.
\onlinecite{emrspaper}).

Comparing the results of our theoretical calculation with the
experimental size dependence of the electron  $g$-factor  we have
determined the surface parameter $Ta_0=-0.6\pm 0.05$ $\AA$ in bare
CdSe nanocrystals. Fitting  the experimental data, we used $E_p=19.0$
eV, that has been obtained  independently from bulk
measurements\cite{seisyan} and that  describes better the
experimental data than   $E_p=17.5$ eV, previously used for CdSe in
Refs. \onlinecite{josa,norris}. The surface parameter $Ta_0$
characterizes the electronic properties of the surface and should be
considered as the additional one to the set of parameters that
describes the bulk properties of semiconductors. In CdSe nanocrystals
prepared by a different technique $Ta_0$ can be different.

The extracted absolute value and the negative sign of $Ta_0=-0.6$
$\AA$ is consistent with our theoretical expectations for $Ta_0$ in
studied CdSe nanocrystals. Its value is very close to the theoretical
value of the surface parameter for semiconductors with a symmetrical
band structure\cite{bc} $|(Ta_0)_s|\equiv
a^*=\sqrt{\hbar^2/2E_pm_0}\approx 0.45$ $\AA$. It can also be shown
within the eight band effective mass model, that the negative sign of
the surface parameter does not allow the existence of the surface
localized states with $E<0$ (gap states) (one can find similar
consideration in Ref. \onlinecite{bc}). Indeed, the CdSe samples
studied in Refs. \onlinecite{gupta,gupta02} show very high PL quantum
efficiency  and do not show  deep gap transitions. The negative
parameter $Ta_0$   leads also to an additional nonparabolicity,
bowing the size dependence of the electron energy levels (see Fig.
\ref{levels}) and therefore may describe the unexplained experimental
size dependence of the 1S electron level in small CdSe
nanocrystals.\cite{norris}

Using the GBCs we have found also the direct surface contribution to
the spin--orbit effects in zero and weak external magnetic fields.
The surface contribution to the zero field spin--orbit splitting is
similar to the interface contribution obtained in Refs.
\onlinecite{bassani,vasko} for $2D$ electrons in planar quantum wells
by using the spin--dependent boundary conditions. It has been pointed
out in Ref. \onlinecite{bassani} that the interface contribution to
the Rashba spin--orbit term in the  $2D$  Hamiltonian\cite{rashba} is
related to discontinuity of the band parameters at the semiconductor
heterointerface and that this contribution  is an additional one to
those connected with the space charge and/or the external electric
field. The importance of the effects described by the Rashba term for
the 2D electrons confined near the curved surface\cite{entin,romanov}
and cylindrical semiconductor quantum dots\cite{vosk} has been
emphasized recently.  The Rashba spin--orbit term in our spherical
dots  is a direct consequence of the GBCs for the envelope function.
The same consideration can be made for cylindrical dots or any other
nanostructure geometry (the results will be published elsewhere). The
GBCs provide an important connection between the constant that
describes the magnitude of the  spin--orbit  term and surface
conditions in nanostructures.

{\em In conclusion}, we have shown an important influence of the
semiconductor surface on the electron energy structure in bare
spherical nanocrystals. The effect of the surface has been modeled
through the choice of the boundary condition parameter that describes
the nonzero value of the envelope function at the nanocrystal
surface. The additional nonparabolicity of the quantum size energy
levels, the  spin--orbit splitting of the electron quantum size
levels, and the additional magnetic moment of the electrons have been
shown to be induced by the surface.  The effects are significant in
small nanocrystals and their considerations require a multiband
effective mass approach because interband coupling is important
there. The analysis of the experimental data allows us to determine
the appropriate parameter of the boundary conditions that
characterize the surface in studied bare CdSe semiconductor
nanocrystals.

\begin{acknowledgments}
 The authors thank I. A. Merkulov and B. K. Meyer for helpful
discussions and J. Tischler for a critical reading of the manuscript.
The work of A. V.  Rodina was supported in part by the Alexander von
Humboldt Foundation, Deutsche Forschungsgemeinschaft (DFG) and the
Swiss National Science Foundation.  Al. L. Efros thanks the U.S.
Office of Naval Research (ONR), U.S. Department of Energy (DOE), and
the DARPA/QuIST program for financial support.  A. Yu. Alekseev
acknowledges support from the Swiss National Science Foundation and
of the grant of INTAS 99-1705.
\end{acknowledgments}

\appendix*

\section{Effect of the small surface perturbation on the
 quantum size level energy in spherical nanocrystals}

We are interested in deriving a variation of the electron level
energy  $E$ caused by a small variation of the parameter $\varsigma $
that characterizes the surface BC for the bulk wave function
$f(r,\varsigma )$. The GBCs at  the spherical surface  of the
nanocrystal with radius $r=a$ can be written as
\begin{equation}
f^{'}(a,\varsigma )=f(a,\varsigma ) A(\varsigma ) \, ,
\label{BCsig}
\end{equation}
where $A(\varsigma )$ is the real number constant. The function
$f(r,\varsigma )$ is the solution of bulk Schr\"{o}dinger equation $
D \hat k^2 f(r,\varsigma ) = E(\varsigma ) f(r,\varsigma )\,$ where
the constant $D$ is independent of $\varsigma $. Taking a derivative
of the Schr\"{o}dinger equation on $\varsigma$, multiplying both
parts of the resulting equation by   $f^{*}$ and integrating it over
the sphere volume, one can obtain:
\begin{equation}
\frac{\partial E}{\partial \varsigma } = D a^2 \left. \left(
\frac{\partial f}{\partial \varsigma }f^{'*} -\frac{\partial
f^{'}}{\partial \varsigma }f^{*}  \right)\right|_{r=a} \, .
\label{fluxsig}
\end{equation}
Taking a derivative of Eq. (\ref{BCsig}) and substituting $\partial
f^{'}(a,\varsigma )/\partial \varsigma  = \partial A(\varsigma
)/\partial \varsigma  f(a,\varsigma ) + A(\varsigma ) \partial
f(a,\varsigma )/\partial \varsigma $ into Eq. (\ref{fluxsig}) one
arrives at  the final expression for the energy variation:
\begin{equation}
\frac{\partial E}{\partial \varsigma } = - D \,\frac{\partial
A(\varsigma )}{\partial \varsigma}\, | f(a)|^2 a^2 \, . \label{varE}
\end{equation}
Substituting $D= \hbar^2/2m_c(E_l)$ and  $A(\varsigma ) =
m_c(E_l)/m_0 \left[1/(Ta_0) + \varsigma \right]$ with $\varsigma = -
\delta k_l^\pm(E_l)$ into this equation we obtain Eq. (\ref{sosur})
for the energy correction $\Delta E_l^{\pm}$  to the averaged energy
of the electron level with the orbital momentum $l$  caused by the
small perturbation of the boundary conditions of Eq. (\ref{bcfso}).

\begin{thebibliography}{100}
\bibitem{errow} In Ref. \onlinecite{emrspaper} we have   mistakenly associated this  term  with the surface contribution to the orbital Zeeman effect.
\end{thebibliography}
\end{document}